\documentclass[a4paper,11pt]{article}
\usepackage{jheppub} 
\usepackage{lineno}

\usepackage{graphicx}
\usepackage{color}
\usepackage{amsmath}

\title{Topological hint to the information paradox and firewall concept for black holes}

\author{J. E. Jacak}
\affiliation{Wroc{\l}aw University of Science and Technology,\\
Wybrze\.ze Wyspia\'nskiego 27, 50-70 Wroc{\l}aw, Poland}

\emailAdd{janusz.jacak@pwr.edu.pl}

\abstract{We demonstrate that at the rim of photon sphere of a black hole the quantum collective transition takes place in any multiparticle system of indistinguishable particles at  passing inward this limiting sphere. This transition is related to the change of the homotopy class of particle trajectories at this sphere causing the local decay of quantum statistics associated with the emission of radiation taking away the energy and entropy of the infalling matter. The matter bereft of its former  entropy  smoothly passes next (in the proper time) the event horizon without any violation of the Einstein equivalence principle nor of the quantum unitarity. No firewall is required on the event horizon because the information does not get lost in the black hole, having previously escaped into outer space at the distance of one and a half of Schwarzschild radius from the black hole central singularity.}


\begin{document}
\maketitle
\flushbottom

\section{Introduction}
Since the illuminating papers by Hawking \cite{hawking} and Bekenstein \cite{bekenstein}, who pointed on the entropy behaviour of black holes in view of second law of thermodynamics, the problem of the fate of an information encoded in matter falling into a black hole is still open. The proposed temperature of black holes and the Hawking-Unruh radiation \cite{hawking,unruh} appeared as a breakthrough quantum property of black holes. However, it has been proved \cite{hawking1} that this radiation is fully random, and cannot carry out the information specific to the matter consumed by the black hole, which causes the information paradox. Trials to solve it via the quantum entanglement of particle-antiparticle pairs at the event horizon and the escape of a particle associated by the falling to the inside of the black hole of its antiparticle partner, also did not solve the paradox especially in view of entanglement property called as its monogamy. To solve the arisen paradox the concept of the firewall on the event horizon was proposed \cite{polczinski}. Polchinski with co-authors \cite{polczinski} stated that the paradox may eventually force to give up one of three time-tested principles: Einstein's equivalence principle, unitarity, or existing quantum field theory. The firewall concept is one of  possible information paradox  solutions. The Hawking-Unruh radiation has not been observed as of yet and the firewall idea is still controversial as it breaks Einstein's equivalence principle \cite{polczinski}. Moreover, the holographic formulation for quantum gravitation by Maldacena \cite{maldac} strengthened the belief that the information should be conserved during the falling of the matter into black holes in accordance to unitary quantum evolution. Thus the fundamental premises of quantum mechanics and of gravitation remain still in conflict in the case of black hole quantum behaviour, which poses a  significant problem for future quantum gravitation theory \cite{polczinski}.

Hawking radiation \cite{hawking} has been resolved to the entanglement of two particles -- escaping and falling into a black hole. Simultaneously, new Hawking radiation must be entangled with the old Hawking radiation, which leads to a conflict with the principle called  "monogamy of entanglement". To avoid this problem, the entanglement  between the infalling particle and the outgoing particle must somehow get immediately broken. Breaking this entanglement would release large amounts of energy, thus creating a searing black hole firewall at the black hole event horizon. Such a solution causes, however, a violation of Einstein's equivalence principle, which states that the free-falling is indistinguishable from floating in an empty space. No entanglement between the emitted particle and previous Hawking radiation would require, on the other hand, black hole information loss, a controversial violation of unitarity.

In a merger of two black holes, the characteristics of a firewall (if any) may leave a mark on the outgoing gravitational radiation as echoes when waves bounce in the vicinity of the fuzzy event horizon. The expected quantity of such echoes is theoretically unclear, as no physical model of  firewall is provided. Afshordi and others argued there were tentative signs of some  such echo in the data from the first black hole merger detected by LIGO \cite{23}, but more recent work has argued there is no statistically significant evidence for such echoes in the data \cite{24}.

In the present paper we contribute to the related situation with the new finding that close to the event horizon of a black hole the homotopy-class of particle trajectories changes qualitatively and this topological universal property causes the decay of quantum statistics in any multi-particle system of indistinguishable particles passing inward  the rim of the photon sphere (in the distance of $1.5 r_s$ from the central singularity of a black hole in Schwarzschild metric, where $r_s=\frac{2GM}{c^2}$ is the event horizon radius, $G$ is the gravitational constant, $M$ is the mass of a black hole, $c$ is the light velocity in vacuum). In particular, all fermionic systems structured according to Pauli exclusion principle decay at passing the photon sphere rim as beneath this limiting sphere  none quantum statistics can be assigned to identical indistinguishable particles \cite{jcap,pra2023}. This quantum collective transition is accompanied with the release of the energy accumulated in Fermi spheres of fermions and of all other energy related to structured by Pauli principle multi-particle systems like multielectron atoms, ions or molecules. Such systems bereft of Pauli exclusion principle constraint, lose their internal structure. In particular, all multielectron atoms transform into heavy hydrogen-type ones, rejecting all chemical compounds and ordered states. All of these are suddenly devoid of information. 
We present this effect in  more detail and point out its significance for the mentioned above problem with the information and entropy behaviour of matter falling into black holes.

\section{Result}
\subsection{Change of the homotopy class of particle trajectories at passing the photon sphere rim}

The destruction of quantum statistics and the related release of the energy accumulated in multi-particle systems, when they are passing the photon sphere rim, considerably changes the situation for discussing of the information paradox or the concept of a firewall at the event horizon of a black hole. The information encoded in  structured matter is related closely with quantum statistics, which for fermionic elementary particles is responsible for Pauli exclusion principle restriction important for any structured matter form. Hadrons and nuclei, next multielectron atoms, molecules and any other macroscopic multi-particle systems must preserve this rule, which from the bottom up to the macroscopic level shapes these systems, as it is not allowed for identical indistinguishable fermions to occupy simultaneously the same quantum state. In particular, indistinguishable fermions cannot approach a place in space  already occupied by another. This leads to a quantum degeneracy repulsion not related with any fundamental force, but exerting a giant strength. The hard core repulsion of hadrons is the result of Pauli exclusion principle for quarks building protons and neutrons and it overcomes their strong attraction and prevents the collapse of atom nuclei. The quantum degeneracy pressure prevents also the collapse of white dwarfs (below the Chandrasekhar limit \cite{chandrasekhar}) or of neutron stars (below the Tollman-Oppenheimer-Volkoff (TOV) limit \cite{tolman,volkoff,olandau}). The same concerns multielectron atoms and molecules giving rise to the whole chemistry of condensed matter. Also homogeneous multi-particle Fermi systems must preserve this rule in collective states, like free electrons in metals or electrons and protons in astrophysical accretion discs \cite{jcap,jhep2022}. Any fermionic structured matter accumulates enormously large energy, because fermions can occupy the quantum stationary single-particle states only one after the other, and a ladder of occupied states is as large as the number of fermions in the system. However, this energy could be released in the event that the Pauli exclusion principle would be locally abrogated.

Note that the Pauli exclusion principle for fermions leads to the entanglement of even  free particles, as the Stater determinant-form antisymmetric multi-particle wave function for indistinguishable fermions is non-separable. Similar entanglement occurs for indistinguishable bosons -- their symmetrized multi-particle wave functions are also non-separable in a  productive tensor  Hilbert space for the  whole system. At approaching the event horizon of a black hole, when passing the photon sphere rim (in the distance $1.5 r_s$, $r_s=\frac{2GM}{c^2}$, from the Schwarzschild black hole singularity) the quantum statistics is locally waived off due to the topological reason \cite{pra2023,jhep2022}. All the energy and entropy/information accumulated due to the entanglement is being released.

In the case of an uncharged and non-rotating black hole, the space-time folded by the general-relativistic gravitation singularity can be described by the solution of Einstein equations in the form of Schwarzschild metric \cite{schwarzschild},
\begin{equation}
	\label{metryka1}
	\begin{array}{l}
		-c^2d\tau^2=-\left(1-\frac{r_s}{r}\right)	c^2dt^2+\left(1-\frac{r_s}{r}\right)^{-1}dr^2
		+r^2(d\theta^2+sin^2\theta d\phi^2),\\
	\end{array}
\end{equation}
where $M$ is the mass of point-like classical gravitational singularity at  $r=0$, $r_s=\frac{2GM}{c^2}$ is the Schwarzschild radius -- the radius of the event horizon of a black hole defining the region, from which neither matter nor radiation can escape. In (\ref{metryka1}) $\tau$ is the proper time and $t, r, \theta, \phi$ are non-folded time-space coordinates (in spherical variables), the same as for a remote observer. In the metric (\ref{metryka1}) two singularities occur -- the first one at $r=0$ is the essential singularity related to the gravitational centre and the second one (due to the second term in r.h.s. of Eq. (\ref{metryka1}) at the event horizon. However, the latter singularity is  apparent, called as the coordinate singularity and can be removed by  the change to another coordinate system, for example Lemaitre, Eddington–Finkelstein, Kruskal–Szekeres, Novikov or Gullstrand–Painlevé coordinates \cite{novikov,kruskal,szekeres}. The change of the coordinate system corresponds to different slicing of the same gravitationally folded space-time into its spatial and temporal components. The reason of the apparent singularity at the event horizon is related to the fact that it does not exist a stationary (time-independent) metric, which could describe simultaneously inner and outer regions with respect to the event horizon, and the stationary Schwarzschild metric (\ref{metryka1}) properly describes only the outside region. If one changes to e.g., Kruskal-Szekeres \cite{kruskal,szekeres} or Novikov \cite{novikov} non-stationary metrics, the singularity at the event horizon disappears and particles can smoothly pass the event horizon within the finite length of the proper time, and terminate any  their movement in the central singularity, also within a finite length of the proper time, despite  in Schwarzschild metric (\ref{metryka1}) the falling of  particles onto the event horizon   takes  infinite time $t$ for  a remote observer and the particles never cross the event horizon for such an observer. In Schwarzschild metric the inner volume of the event horizon sphere is zero, though is non-zero in other metrics \cite{novikov,kruskal,szekeres}. Despite these differences between  metrics for the same folded space-time, the central singularity and  event horizon are universally defined in all  metrics. Of particular importance is the Kuskal-Szekeres metric \cite{kruskal,szekeres}, which is the complete solution of the Einstein equations for point-like mass $M$, i.e., is analytic in the whole range of its definition. Schwarzschild metric (\ref{metryka1}) is, however, very convenient for the trajectory analysis, as the stationary space coordinates are rigid and the same as for a remote observer. Especially easy and transparent is  demonstration  in this coordinates of the trajectory homotopy class change at passing the photon sphere rim at $r=1.5r_s$. Note, that the classes of homotopy are invariant to changes in coordinate systems.

The homotopy class of particle classical trajectories for multi-particle systems of identical particles is of fundamental significance for the quantum statistics assigned to these particles. Quantum statistics is topologically conditioned \cite{leinaas1977,sud,lwitt-1}, and displays the symmetry properties of multi-particle wave functions at interchanges of their arguments – the classical positions of particles on some manifold. Due to quantum indistinguishability of identical particles, the probability defined by the quantum mechanical wave function of a specific positioning of particles is immune to particle exchanges, but the overall phase of the wave function must change according to a scalar unitary representation of the braid describing a particular renumbering of particles on some manifold. This braid is a multi-strand bundle of individual classical trajectories of all particles displaying their exchange in the multi-particle configuration space $F_N=(M^N-\Delta)/S_N$ (where $M$ is the manifold on which all particles are located, $M^N=M\times M\times \dots \times M$ is $N$-fold product of $M$, to account for all $N$ particles, $\Delta$ is the diagonal subset of $M^N$, subtracted to assure the particle number conservation in the case when at least two particles have the same position, $S_N$ is the permutation group of $N$ elements, and the quotient structure of $F_N$ by $S_N$ accounts for the indistinguishability of identical particles). The first homotopy group $\pi_1({\cal{A}})$ \cite{mermin1979,spanier1966}, i.e., the group of non homotopic classes of closed paths in the space ${\cal{A}}$ (non homotopic means that paths from one class cannot be deformed in a continuous manner without any cutting into a path from other class) is especially convenient to describe and classify particle exchanges. In the case of $F_N$ space, the group $\pi_1(F_N)$ displays all possible renumbering of particles, as closed multi-strand trajectory loops in this case join different particle distributions on a manifold $M$ which differ only by particle numbering (such various distributions of particles on $M$ are unified into a single point in $F_N$, thus multi-strand trajectories joining these distributions are closed loops from $\pi_1(F_N)$). The homotopy group $\pi_1(F_N)$ is conventionally called as the full braid group \cite{birman,mermin1979} and loops from this group are called the braids.

 The first homotopy group $\pi_1({\cal{A}})$ displays the difference between simply- and multiply-connectivity of arc-connected space ${\cal{A}}$ \cite{spanier1966}. In the latter case the braid group $\pi_1({\cal{A}})$ is non-trivial, whereas in the former is trivial (i.e., has only a neutral element). $F_N$ space is usually multiply-connected and $\pi_1(F_N)$ is non trivial except of some special case we will describe below. The full braid group $\pi_1(F_N)$ depends on the topology and the dimension of $M$. For $dim M>2$ the full braid group is always a simple permutation group, $\pi_1(F_N)=S_N$, i.e., is the finite ($dimS_N=N!$) group \cite{birman,mermin1979}. For $dim M=2$ the full braid group  is the infinite Artin group \cite{artin1947}, which is multicyclic countable group generated by the generators $\sigma_i$, $i=1,\dots, N-1$, describing elementary exchanges of $i$-th particle with $(i+1)$-th one, when other particles stay in rest. Though the similar generators  generate also the permutation group $S_N$, but the generators of Artin group fulfil different conditions than those for the permutation group. For Artin group these relations are as follows \cite{birman},
\begin{equation}
	\label{relacje}
\begin{array}{l}
\sigma_i \sigma_{i+1}\sigma_i=\sigma_{i+1}\sigma_{i} \sigma_{i+1}, \;\mbox{for} \;1\leq i\leq N-2,\\
\sigma_i \sigma_j=\sigma_j\sigma_i, \;\mbox{for} \;1\leq i,j\leq N-1,\; |i-j|\geq 2,\\
\end{array} 
\end{equation}
whereas for the permutation group $\sigma_i^2=e$ ($e$ is the neutral element in the group, for the  Artin group $\sigma_i^2\neq e$).
 Except of the full braid group it is also defined the pure braid group $\pi_1(P_N)$, where $P_N=(M^N-\Delta)$, i.e., $P_N$ is the multi-particle configuration space of distinguishable $N$ identical particles; the pure braid group is a subgroup of the full braid group and is of significance for mathematical homotopy group structure, but not for quantum physics \cite{birman,mermin1979}.

The permutation group $S_N$ has only two scalar unitary representations, defined on generators $\sigma_i\rightarrow \left\{\begin{array}{l} e^{i0},\\
	e^{i\pi},\\ 
	\end{array}\right.$ which assign quantum statistics of bosons and fermions, respectively. The Artin group has infinite number of different quantum statistics assigned by scalar unitary representations $\sigma_i\rightarrow e^{i\alpha}$, $\alpha \in[0, 2\pi)$ corresponding to exotic quantum particles anyons (including bosons for $\alpha =0$ and fermions for $\alpha=\pi$) \cite{wilczek}. Anyons are related to 2D manifolds and display statistics of quasiparticles and quasiholes for FQHE states (Laughlin states) \cite{laughlin2,prange}.

In the present paper we show another exotic case for identical indistinguishable particles, which results in homotopy constraints imposed onto particle trajectories close to the event horizon of a black hole. 
We show that any quantum statistics cannot be assigned to a system of identical indistinguishable particles (on $M$ of arbitrary dimension) if it passes inward the rim of the photon sphere of a Schwarzschild black hole. The photon sphere rim has the radius of $1.5 r_s$ (for non-rotating, uncharged black hole), which is the radius of the innermost unstable circular orbit for any particle including photons. The stable circular orbits terminate for massive particles at $r=3r_s$. Below the sphere with radius $3r_s$ up to $1.5r_s$ only unstable circular orbits are possible. Below the photon sphere rim at $r=1.5r_s$ any trajectory in the form of a closed loop for particles does not exist. If particles (including photons) pass the photon sphere rim inward, they unavoidably spiral onto the event horizon. Spirals are short, the shorter the larger is energy at some angular momentum of a particle, though for a remote observer the fall onto the event horizon along these short spirals takes infinite time. For such an observer particles never touch the event horizon, However, if to switch to the proper time (of the observer attached to a falling particle), then particles smoothly pass the event horizon within a finite proper time period, and next, also along a spirals, focus in the central singularity during a subsequent small finite proper time period.

\begin{figure}
	\centering
	\includegraphics[width=0.8\columnwidth]{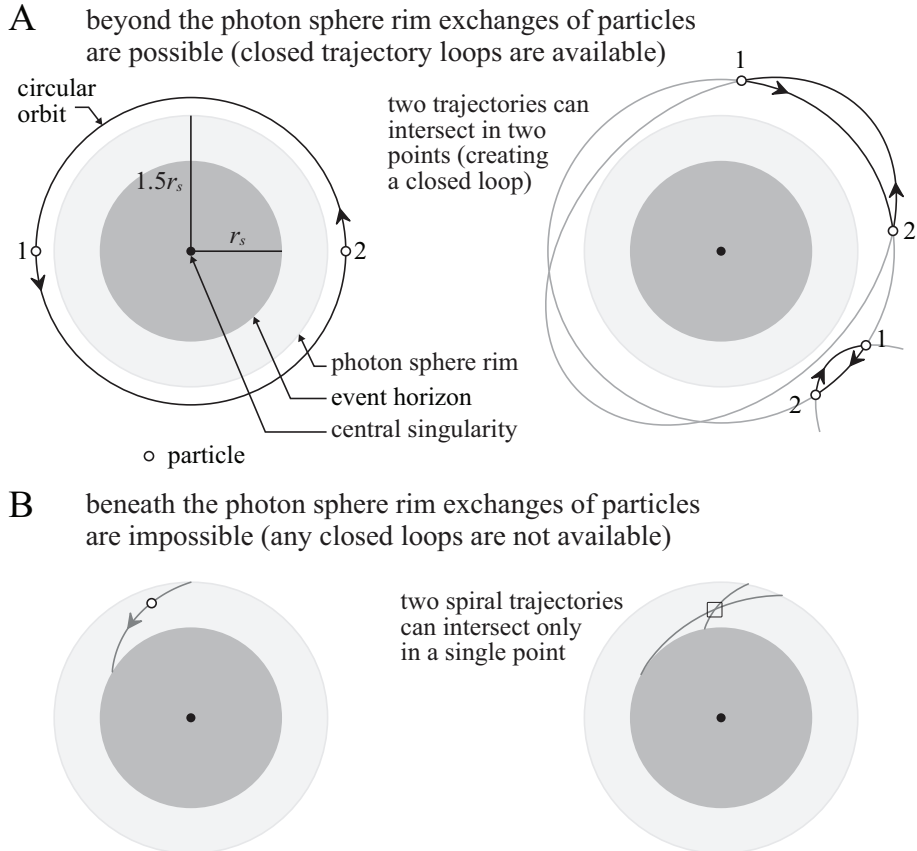}
	\caption{\label{gc} Simplified pictorial illustration of the change of trajectory homotopy at passing the innermost unstable circular orbit of a black hole. If conic section-type trajectories are available (beyond the photon sphere rim) then the particle position interchanges are possible, like the exchange of particles 1 and 2 along a circular orbit. Conic section-type trajectories can intersect in two points and local, arbitrarily small, closed loops can be constructed from pieces of such trajectories (upper picture). When only short spiral trajectories are admitted beneath the photon sphere rim and particles unavoidably fall towards the event horizon (lower picture), then particles cannot mutually interchange their positions because these spirals can intersect in only one point and do not create loops needed for particle exchanges.}
\end{figure}

The qualitative change of the trajectory homotopy takes place at the photon sphere rim in the distance $1.5 r_s$ from the central singularity. The photon sphere rim in the metric (\ref{metryka1}) is defined by the innermost unstable circular orbit for any particle (cf. Methods \ref{aaa}). Any massive or massless particle unavoidable spirals onto the event horizon if it passes the photon sphere rim in inward direction. Beneath the photon sphere rim none circular or other closed orbits are possible. No local of arbitrary small size closed loops built of trajectories exist here. This is different compared to Newton gravitation centre, for which circular orbits and local arbitrarily small loops are available at arbitrary close vicinity to the point-like gravitation centre. For Newton gravitation centre various conic section trajectories are accessible for particles arbitrarily close to the centre and these conic sections can cross at two arbitrarily close points making possible to create closed small local loops built of trajectories. Even though for large distance from the gravitation centre the Schwarzschild trajectories can be approximated by conic sections (modified by some additional deformation, e.g., a precession of elliptical orbits, like that observed for Mercury in Sun gravitation), in close vicinity of the event horizon conic section-like trajectories completely disappear below the photon sphere rim. Conic sections allow the formation of closed arbitrarily small local loops from pieces of various trajectories because conic sections can cross in two points (like circle with ellipse, hyperbole or parabola), as illustrated schematically in Fig. \ref{gc}. This is in contrary to short spirals below the innermost unstable circular orbit, where the accessible spirals (described by Eqs (\ref{promien}) and (\ref{phase}) in Methods \ref{aaa}) can intersect  in a single point only. The innermost unstable circular orbit thus separates  two different space regions with different classes of homotopy of trajectories in systems of identical indistinguishable particles.

\begin{figure}
	\centering
	\includegraphics[width=0.8\columnwidth]{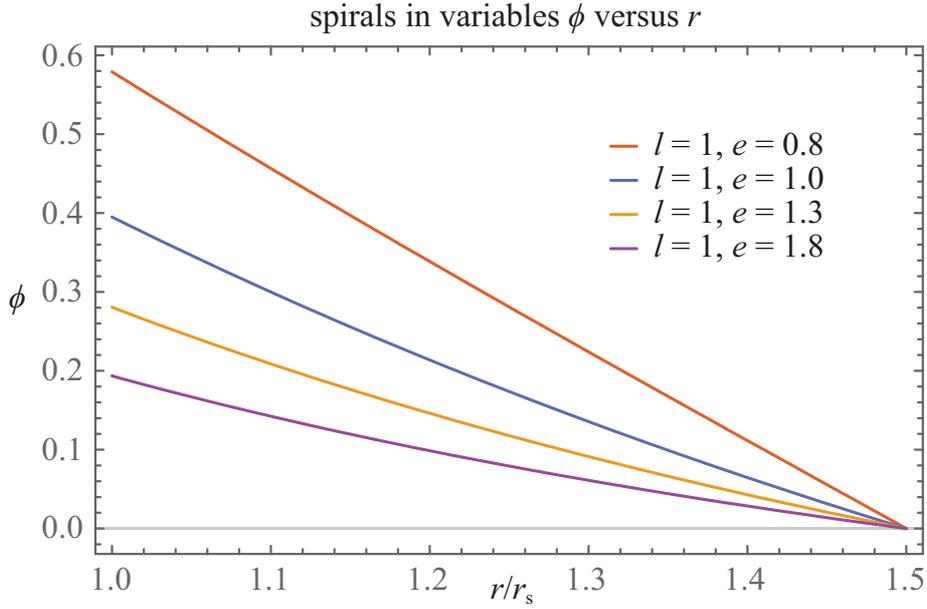}
	\caption{\label{trajektorie-bh} Spiral trajectories for particles passing the photon sphere rim of a black hole, shown in spherical variables, phase shift $\phi$ versus radius $r$ -- the solutions of Eq. (\ref{phase}) for selected motion constants ${\cal{L}}$ and ${\cal{E}}_0$ (in the figure legend, $l=\frac{\cal{L}}{mcr_s}$, $e=\frac{{\cal{E}}_0}{mc^2}$). }
\end{figure}

\begin{figure}
	\centering
	\includegraphics[width=0.8\columnwidth]{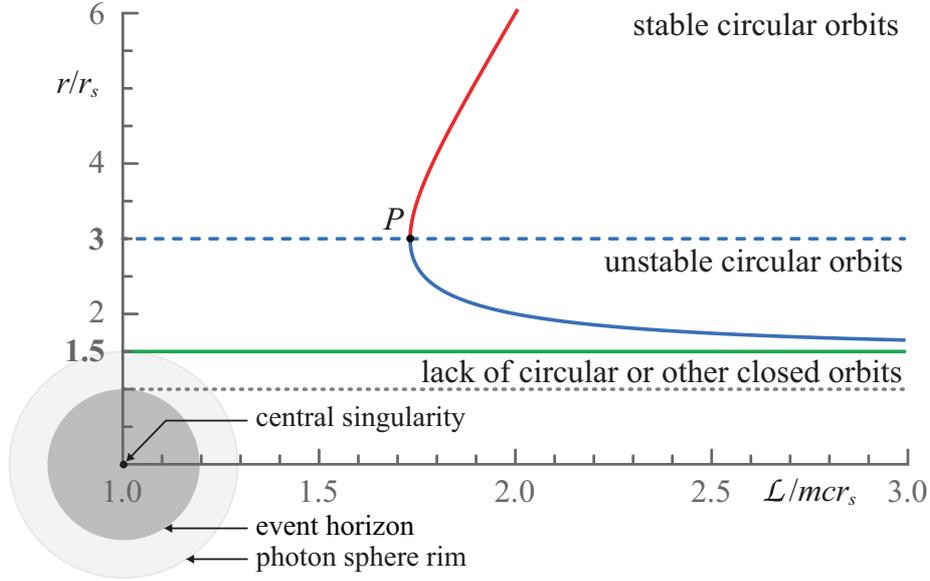}
	\caption{\label{photosphere} Radii of stable (red) and unstable (blue) circular orbits in Schwarzschild geometry. The innermost circular stable orbit occurs at $r=3r_s$ -- point $P$ on the level of dashed line in blue colour (the coordinates of point $P$ are, angular momentum ${\cal{L}}=\sqrt{3}mcr_s$ and energy ${\cal{E}}_0=\sqrt{\frac{8}{9}}mc^2$, $m$ is the mass of a particle), whereas the innermost unstable circular orbit occurs at $r=1.5 r_s$ (asymptotic dotted line in green colour, ${\cal{L}}\rightarrow \infty$ and ${\cal{E}}_0\rightarrow \infty$). Below $r=1.5r_s$ neither circular nor other closed orbits exist (cf. Methods \ref{aaa}). The innermost unstable circular orbit defines the photon sphere rim in Schwarzschild geometry. 	}
\end{figure}

Homotopy of trajectories (analysed in Methods \ref{aaa}) can be summarised in Fig. \ref{photosphere} -- the upper curve (red) gives positions of stable circular orbits of a particle in the metric (\ref{metryka1}) (with respect to angular momentum of the particle ${\cal{L}}$, the motion integral and the lower curve (blue) gives positions of unstable circular orbits. 
The upper curve terminates in the point $P$ at $r=3r_s$. This point defines the innermost stable circular orbit. 
The position of the innermost unstable circular orbit is at $r=1.5 r_s$ for ${\cal{L}} \rightarrow \infty $. The orbit $r=1.5 r_s$ is also the unstable circular orbit for photons (by taking the limit $m=0$ for a particle mass), thus it defines the photon sphere rim in Schwarzschild metric. 

Below the innermost unstable circular orbit the dominating term in the effective potential (Methods, Eqs (\ref{differential}) and (\ref{potencjal})) is $\sim -\frac{{\cal{L}}^2}{r^3}$, which causes an unavoidable spiral movement towards the event horizon of any particle despite its energy and angular momentum and local mutual interaction between particles. The phase shift for such spirals is all the more limited the larger ${\cal{E}}_0$ for given ${\cal{L}}$ (cf. Fig. \ref{trajektorie-bh}), which magnitude growth enhances the attraction to the centre  $\sim{\cal{L}}^2$, on the other hand. From pieces of these short spirals it is impossible to form any closed local loop required for local interchanges of particles. Note that braids for particle interchanges must be built of real existing trajectories -- in the case when such trajectories are not available, then braids (and quantum statistics) do not exist -- for a proof cf. Methods \ref{ggg1}.

The absence of closed trajectory loops for particles beneath the photon sphere rim precludes the possibility of mutual interchanges of particle positions in any systems of $N$ identical indistinguishable particles passing this rim \cite{pra2023}. It is unreachable to form a closed loop from pieces of spirals defined by Eqs (\ref{promien}) and (\ref{phase}) in contrast to conic-section-like trajectories outside of the photon sphere. Hence, for $r\geq 1.5r_s$ the fermionic or bosonic quantum statistics can be assigned \cite{mermin1979,birman} in contrary to the region $r\in(r_s,1.5r_s)$, where none quantum statistics is defined \cite{pra2023}.

\subsection{Decay of quantum statistics at the photon sphere rim}

Quantum statistics can be assign to a multiparticle system of identical indistinguishable particles, for which the classical configuration space has the form, $F_N=(M^N-\Delta)/S_N$, where $M$ is a manifold on which all particles are placed, $M^N=M\times\dots\times M$ is $N$-fold product of $M$ to account all particles (assuming that the whole $M$ is accessible to each particle equally). $\Delta$ is the diagonal subset of $M^N$ collecting points with at least two particle coordinates coinciding, and removed in order to preserve particle number conservation -- for more rigorous mathematical background cf. Methods \ref{ggg1}. The indistinguishability of particles is introduced by the quotient structure by the permutation group $S_N$ in the space $F_N$ definition -- points which differ only by particle numbering are unified in $F_N$. The space $F_N$ is usually multiply-connected and its fundamental group (the first homotopy group \cite{spanier1966,mermin1979}) $\pi_1(F_N)$ is non-trivial. The explicit form of $\pi_1(F_N)$, conventionally called as the braid group, depends on the dimension of $M$ manifold. For 3D $M$, $\pi_1(F_N)=S_N$, i.e., is the finite permutation group. However, for 2D $M$, $\pi_1(F_N)$ is infinite and for $M=R^2$ (the plane) the braid group is the infinite Artin group $B_N$ \cite{artin1947,birman}. Braid groups are multicyclic groups generated by a finite set of generators $\sigma_i$, $i=1,\dots, N-1$ -- the exchanges of $i$-th and $(i+1)$-th particles at arbitrary but fixed particle numbering. Hence, the braid groups are countable.

Quantum statistics is defined by the scalar unitary representation (1DUR {\it one dimensional unitary representation}) of the braid group for a particular system \cite{lwitt-1,mermin1979}. As the permutation group (which is a braid group for any 3D $M$) has only two different 1DURs defined on generators as, $\sigma_i\rightarrow e^{i0}\;\text{or}\;e^{i\pi}$, then in 3D case only bosonic or fermionic statistics are possible, respectively. For  Artin group  infinite number of different 1DURs exist, $\sigma_i\rightarrow e^{i\alpha}$, $\alpha\in[0,2\pi)$ leading to anyonic statistics \cite{wilczek}. In the case when the space $F_N$ is simply-connected, then its fundamental group $\pi_1(F_N)=\{e\}$ is trivial (here $e$ denotes the neutral group element), and any quantum statistics cannot be assigned, because $e\cdot e = e$ and 1DUR$(e)=1$ (this is not bosonic statistics, as $e$ is not exchange of particles).

Beneath the rim of the photon sphere of a Schwarzschild black hole, the space $F_N$ of any multi-particle system of indistinguishable identical particles is simply-connected, because any trajectories for particle exchange do not exist there. An exchange of two particles needs the existence of a non-trivial trajectory loop in $F_N$ braiding these two selected particles (cf. Methods \ref{ggg1}). Because the fundamental group $\pi_1$ is the collection of closed loop classes, which cannot be continuously deformed without cutting, one onto another one, between different classes (which are non homotopic \cite{spanier1966}) and in $F_N$ various distributions of particles which differ only by particle numbering are unified into a single point, the closed loops from $\pi_1(F_N)$ define particle position exchanges and the $N$-strand trajectories, creating these loops, are closed braids. The existence of the elementary braid -- the exchange of only two particles when other stay at rest, is equivalent to the existence of a closed loop in $M$ built of two different single-particle trajectories joining two particles with different positions on $M$, i.e., to the existence of at least two  distinct single-particle trajectories in $M$ intersecting in these two points. However, single-particle trajectories for any $M$ located below the photon sphere rim cannot intersect in two points, thus the required closed loops cannot be implemented there. This is in contrast to the upper region outside the photon sphere, where trajectories of conic-section type can intersect in two points, similarly as pure conic-section trajectories in the whole unfolded space with Newton gravitation centre. In the case of general-relativistic gravitational singularity, the space $F_N$ is multiply-connected beyond the photon sphere rim, but is simply-connected beneath this limiting sphere. Hence the braid group of any $N$-particle system of indistinguishable particles is $S_N$ outside of the photon sphere (as in 3D position space even gravitationally folded) and $\{e\}$ beneath the photon sphere rim, where $e$ is a neutral element of the group. In the latter region quantum statistics cannot be assigned. In this region the Pauli exclusion principle is locally waived off, in particular.

The local decay of quantum statistics is associated with the release of huge energy at passing inward the photon sphere rim by any multi-particle macroscopic system of fermions structured due to Pauli exclusion principle. This energy emission in the form of e-m radiation according to the Fermi golden rule for admitted there quantum transitions, takes away the entropy and the particles devoid of quantum statistics create a pure quantum state of individual particles which cannot interchange their positions, like in an ideal crystal invoked in third law of thermodynamics by Nernst. The matter that next crossed (in finite length of proper time) the event horizon does not carry the entropy or information, which neither causes the information paradox nor the firewall is required there. The role of some-kind of  firewall has been taken by the visible to any distant observer the radiation burst released during the passing the photon sphere rim relatively distantly from the event horizon.

Any multi-particle system which passes the photon sphere rim loses its quantum statistics. 
This does not violate the Pauli theorem on spin and statistics as shown in Methods \ref{C}. 
The so-called exclusion principle (also stated by Pauli) asserts that quantum particles of fermionic type cannot share any common single-particle quantum state. Fermions cannot approach a space region already occupied by another fermion, and thus they mutually repulse themselves. This is called as the quantum degeneracy repulsion and the related pressure is the origin of stopping the collapse of white dwarfs or neutron stars. In the former case the degeneracy pressure of electrons plays the role \cite{chandrasekhar}, whereas in the latter case of neutrons \cite{tolman,volkoff,olandau}.

The exclusion principle for fermions leads also to the formation of Fermi sphere in the case of large number of identical fermions located in some volume, when the chemical potential $\mu$ (the energy increase due to the addition of a single particle to a thermodynamic multi-particle system) is much greater than the temperature of the system in energy scale, $k_BT$, $k_B$ is the Boltzmann constant. In such a case, referred to as the quantum degenerate Fermi system, the particles are forced to occupy consecutive  energy single-particle states one by one, resulting in the accumulation of energy. For example, free electrons in a normal metal at room temperature with the typical concentration of order of $10^{23}$ (of order of Avogadro number) per cm$^3$ constitute the Fermi sphere with accumulated energy $\sim 3\times10^{10}$ J/m$^3$. This energy cannot be released in normal conditions because all fermions are blocked in their single-particle stationary states by Pauli exclusion principle, i.e., all lower single-particle states are occupied and thus there is no room for fermions to jump from higher energy states to lower ones.

The situation changes, however, when quantum statistics cannot be defined below the photon sphere rim of a black hole. The energy accumulated in the Fermi sphere of strongly compressed fermion systems can be released here.
The Fermi momentum $p_F$, the radius of the Fermi sphere in the degenerate homogeneous quantum liquid of fermions depends only on particle concentration \cite{ashcroftmermin,abrikosov1975},
\begin{equation}
	\label{fm}
	p_F=\hbar(3\pi^2 \rho)^{1/3},
\end{equation}
where $\hbar=\frac{h}{2\pi}$ is the reduced Planck constant and $\rho=\frac{N}{V}$ is the concentration of $N$ fermions in the spatial volume $V$.
The Fermi momentum is independent of interaction of fermions according to Luttinger theorem \cite{luttinger,abrikosov1975}. This follows from the fact that the phase space volume $V \frac{4}{3} \pi p_F^3$ of $N$ identical particle system corresponds to $n =\frac{V 8\pi p_F^3}{3h^3}$ of single-particle quantum states according to Bohr-Sommerfeld rule \cite{landau1972a} (factor $2$ accounts for doubling of states for $\frac{1}{2}$ spin of fermions). If all these states are filled, i.e., when $n=N$, then one gets (\ref{fm}). The formula for Fermi momentum, as quasi-classically derived by Bohr-Sommerfeld rule (thus, immune to interaction), is independent of interaction of fermions in Fermi liquid. With Fermi momentum the Fermi energy is linked -- the kinetic energy of a particle with Fermi momentum (the chemical potential of fermions at $T=0$ K is equal to the Fermi energy and weakly changes with the temperature growth \cite{abrikosov1975}).

The whole Fermi sphere collects the energy per spatial volume $V$ (neglecting the interaction of fermions),
\begin{equation}
	\label{en}
	\begin{array}{l}
		E=\sum_{\mathbf{p}} \varepsilon(\mathbf{p}) f(\varepsilon(\mathbf{p}))\\
		= \frac{V}{(2\pi \hbar)^3}\int d^3\mathbf{p}\varepsilon(\mathbf{p})f(\varepsilon(\mathbf{p}))\\
		=\int_0^{p_F}dp \int_0^{\pi} d \theta\int_0^{2\pi} d\phi p^2 sin\theta \varepsilon(\mathbf{p}) \frac{V}{(2\pi \hbar)^3}\\
		=\frac{V}{2 \pi^2 \hbar^3}\int_0^{p_F} dp p^2 \varepsilon(p),\\
	\end{array}	
\end{equation}
where the sum runs over occupied states only, what is guaranteed by Fermi-Dirac distribution function $f(\varepsilon(\mathbf{p}))=\frac{1}{e^{(\varepsilon(\mathbf{p})-\mu)/k_BT}+1}\rightarrow_{T\rightarrow 0}1-\Theta(\varepsilon(\mathbf{p})-\varepsilon_F)$ ($\Theta(x)$ is the Heaviside step function and $\varepsilon_F=\varepsilon(p_F)=\mu(T=0)$, $\mu$ is the chemical potential), $p,\theta, \phi$ are spherical variables in momentum space and $\varepsilon(\mathbf{p})$ is the kinetic energy of a fermion equal to $\sqrt{p^2c^2+m^2c^4}-mc^2$ (in relativistic case). The factor $\frac{V}{(2\pi \hbar)^3}$ is the density of quantum states, i.e., the number of single-particle quantum states in the element of the phase space $Vd^3\mathbf{p}$. The energy estimation (\ref{en}) holds also for non-zero temperatures, if $k_BT\ll\mu\simeq\varepsilon_F$ (i.e., when the Fermi liquid is quantumly degenerated).

For a neutron star at TOV limit \cite{tolman,volkoff} with the density of order of $5 \times10^{18}$ kg/m$^3$ (corresponding to 2.3 Sun mass compressed to the compact neutron star with radius of ca. $6$ km), the neutron Fermi sphere energy reaches $0.5 \times 10^{47}$ J, just as the energy of frequently observed cosmic short giant gamma-ray bursts (assuming the isotropy of their sources) -- cf. Table \ref{tab-kw1}. The Fermi energy in this case $\varepsilon_F\simeq 3.4 \times 10^{12}$ K (in $k_B=1$ units), which is much greater than the supposed temperature of the neutron star, of order of $10^6$ K (thus neutrons in a neutron star form a degenerate quantum system).
\begin{table}
	\centering
	\begin{tabular}{l|l|l|l|l}
		\hline 
		$p_F$ [kg m/s] &$ \xi$ [kg/m$^3]$ & $n$& $\varepsilon_F$ [GeV] ([K] if $k_B=1$) & $E$ [J]/luminosity [W] \\ 
		\hline 	
		\multicolumn{5}{c}{unstable neutron star merger of 2.3 $\text{M}_{\odot}$ with radius ca. $6$ km} \\
		\hline 
		$4.67 \times 10^{-19}$ &	$5\times10^{18}$ & $2.4 \times 10^{57}$& ne $0.34$ ($3 \times 10^{12}$) & $4.7 \times 10^{46}$ J\\
		\hline 
		\multicolumn{5}{c}{quasar with BH of $10^9$ $\text{M}_{\odot}$ consuming $0.06$ Earth mass per second ($5.6$ $\text{M}_{\odot}$ per year)}\\
		\hline
		$5.4 \times 10^{-19} $ & $9\times 10^{18}$& $2.14 \times 10^{50}$ &el $1$ ($10^{13}$), pr $0.4$ ($4\times 10^{12}$)&$ 10^{40}$ W\\
		\hline
	\end{tabular}
	\caption{Fermi momentum $p_F$ (acc. to Eq. (\ref{fm})), Fermi energy $\varepsilon_F=\varepsilon(p_F)$ and total energy of the Fermi sphere $E$ released at the collapse of the Fermi sphere, for neutron star (acc. to Eq. (\ref{en})), for quasar luminosity, i.e., energy per second ($\xi$ -- density of compressed electron-hadron matter, $n$ -- total number of neutrons (ne) (electrons (el) or protons (pr))).}
	\label{tab-kw1}
\end{table}

The coincidence of the energy stored in the Fermi sphere of neutrons in a neutron star at TOV limit with the energy of short giant gamma-ray bursts may support the concept that the source (yet unknown) of some of these bursts is a collapse of the Fermi sphere of neutrons when the whole star is compressed to the volume inside its own photon sphere. At the decay of the Fermi sphere of neutrons, the latter liberated from the Pauli exclusion principle constraint fall apart on charged electrons and protons interacting with the electromagnetic field. Rapid allowance of jumping of these particles onto their ground state at the decay of the neutron Fermi sphere, will release a giant flux of isotropic electromagnetic radiation along the Fermi golden rule for quantum transitions (as described in \ref{F}), with dominant component of gamma radiation because of large value of Fermi energy in this case (cf. Table \ref{tab-kw1}).

In another example -- in an accretion disc of a quasar, the density of electron and proton plasma (assuming accretion of neutral hydrogen) grows with the falling of the matter towards the Schwarzschild horizon. The originally diluted neutral gas ionises itself due to friction in the accretion disc and eventually becomes a degenerate Fermi liquid despite the high local temperature. For two component plasma both Fermi spheres of electrons and protons contribute to the energy storage. This energy grows on the cost of the gravitation of a central black hole which compresses particles to high concentration. At the same concentration of electrons and protons (due to the neutrality of plasma in the disc) electrons accumulate larger kinetic energy because of their lower rest mass. 
The energy accumulated in the Fermi spheres of electrons and protons can be released in the form of the electromagnetic radiation if the Pauli exclusion principle is locally waived off at the rim of the photon sphere, due to local decay of quantum statistics. Charged particles couple to the electromagnetic field and the collapse of their Fermi spheres is accompanied with the emission of photons in agreement with Fermi golden rule for quantum transitions between initial states of particles in the Fermi sphere and their ground state.

Super-luminous quasars with central black hole $\sim 10^9$ solar mass consume typically $10$ Sun mass per year ($0.1$ Earth mass per second), when the accretion of the gas is limited only by the uppermost density of matter at the photon sphere rim (similar to the density of neutron star at TOV limit or to the density of atom nuclei). The continuous decay of Fermi spheres of electrons and protons in accretion plasma crossing the photon sphere of a supermassive central black hole releases photons with total energy up to 30 \% of the falling mass \cite{jhep2022,jcap}. This fraction of the mass is the energy (divided by $c^2$) of the Fermi spheres of electrons and protons accumulated on the cost of the gravitation during matter compression in the accretion disc. This rapidly released energy, in the form of radiation, contributes to the luminosity of quasars with ca. $10^{40}$ W from the close vicinity of the event horizon (at passing the rim of the photon sphere at $r=1.5 r_s$) in a better agreement with observations than only radiation from more distant regions of the accretion disc \cite{merloni,6p,24p,shapiro}. If the supply of the matter to the accretion disc is limited by environmental conditions, then the density of particles at the photon sphere rim is not extremal and the efficiency of Fermi sphere decay is much lower than 30 \%. The detailed quantitative estimations, including general gravitation correction to the density are presented in Methods \ref{E}. 
\subsection{Absence of the firewall at the event horizon }
The matter deprived of quantum statistics does not form a classical system of particles but a specific pure quantum state of particles which cannot mutually interchange their positions. For a remote observer the entropy of these particles is zero as well as temperature is zero as for an idealised perfect crystal in the third law of thermodynamics. Only due to the Unruh effect \cite{unruh} the mixed state can be noticed by an accelerating observer (in particular in the proper system of falling particles). The related Hawking-Unruh radiation is completely independent of matter state but is related to the kinematic property in general relativity and the equivalence principle, which still holds at passing of the event horizon. This radiation is completely random \cite{hawking1} and in this way we do not encounter any information paradox at the event horizon. Instead,  at the rim of the photon sphere the radiation is released due to the decay of quantum statistics and local waiving off the Pauli exclusion principle. The radiation is emitted according to the Fermi golden rule in an unitary manner \cite{jcap}. Thus, the entropy of the falling matter is taken away by this radiation before crossing the event horizon. This radiation is of firewall type, but in distinction to that proposed by Polchinski and co-authors \cite{polczinski}, is visible for a remote observer (the Polchinski's firewall must be invisible for a remote observer as it would assist the entanglement breaking at passing the event horizon, which is unobservable for a distant observer). The release of energy accumulated in multi-particle fermion systems does not violate the classical Einstein's equivalence rule, at least for any single particle in an arbitrary pure quantum state, which does not emit any energy at passing the photon sphere rim and conserves its state (with zero entropy as for any quantum pure state).

We observe thus, that the topological effect of the change of homotopy class of particle trajectories at passing the innermost unstable circular orbit of a black hole modifies premises for the discussion of information paradox and related problems. In particular a field theory must be constructed neither for fermions nor for bosons beneath the photon sphere rim.

Disappearance of the quantum degeneracy pressure beneath the photon sphere rim conveniently assists the shrinking of the matter on its way towards the central singularity. This is neither a conventional matter of bosons or fermions nor the classical matter. Instead, it is an accelerating multiparticle system without any statistics of individual particles not allowed to mutually interchange positions (separable pure quantum state), shrinking and moving towards the singular point.

\subsection{Spectrum and timing of short giant gamma-ray burst at unstable neutron star merger collapse}

\label{F}

The mechanism of a collapse of neutron star merger with the mass exceeding the TOV limit \cite{tolman,volkoff} is not known but it is thought that this episode of black hole creation is associated by a short giant gamma-ray burst,  also without knowing the mechanism of conversion of ca. 30 \% of initial merger mass into e-m radiation. Some hints towards understanding of both these phenomena  are supplied by the described decay of quantum statistics in vicinity of the event horizon of a black hole \cite{pra2023,jhep2022}. The rapid waiving off Pauli exclusion principle relieves the internal quantum pressure in the merger allowing its collapse. The simultaneous decay of Fermi spheres of fermions releases a giant accumulated energy in the form of e-m radiation along the Fermi golden rule for quantum transitions of charged particles suddenly admitted at the matter compression beyond the TOV limit (when the whole merger is within its own photon sphere). The energy of the neutron Fermi sphere in the neutron merger at TOV limit largely agrees  with the energy of observed giant short gamma-ray burst of total energy $10^{47}$ J (assuming isotropy of its source). Neutrons in neutron stars are stabilised only by  Pauli exclusion principle, and when the quantum  statistics decays, the neutrons decompose into stable electrons and protons -- charged particles interacting with e-m radiation, for which the Fermi golden rule holds. This rule allows thus for the assessment of kinetics of the Fermi sphere decay and the identification of spectral contents of radiated e-m waves. Both the spectrum of this radiation and its temporal characteristics agree with observations of short gamma-ray bursts.

The probability of quantum transitions per time unit due to interaction of charged quantum particle with e-m wave is given by the Fermi golden rule \cite{landau1972a},
\begin{equation}
	\label{zrf}
	w_{1,2}=\frac{2\pi}{\hbar} \vert\langle 1 \vert\hat{V}(\mathbf{r})\vert 2 \rangle\vert^2\delta(E_1-E_2-\alpha\hbar\omega),
\end{equation}
where $\vert 1(2)\rangle$ is the initial (final) quantum stationary state of a particle with energy $E_{1(2)}$, $\alpha\hbar\omega$ is the energy of emitted photon at the transition $1\rightarrow 2$ (with gravitational redshift close to the black hole horizon, $\alpha=(1-\frac{r_s}{r})^{1/2}$, here for $r=1.5 r_s$). $\hat{V}(\mathbf{r})$ is a quantum operator describing coupling of charged particles to  e-m wave. For relativistic electrons or protons the kinetic energy single-particle Hamiltonian has the form, 
$\hat{H}=\sqrt{(\hat{\mathbf{p}}\mp e\mathbf{A}(\mathbf{r},t))^2c^2+m_{e(p)}^2c^4}-m_{e(p)}c^2$, where $\mathbf{A}(\mathbf{r},t)=\mathbf{A}_0e^{i(\mathbf{q}\cdot\mathbf{r}- cq t)/\hbar}$ is the vector potential of the e-m field for a plane wave (at gauge $div{\mathbf{A}}=0$), $\hat{\mathbf{p}}=-i\hbar\nabla$. Hence, the perturbation  linear with respect to $\mathbf{A}$ attains the form,
\begin{equation}
	\label{zaburzenie}
	\hat{V}(\mathbf{r},t)=\mp\frac{e c^2 \mathbf{A}({\mathbf{r}},t) \cdot\hat{\mathbf{p}}}{\sqrt{\hat{\mathbf{p}}^2c^2+m_{e(p)}^2c^4}}.
\end{equation}	
For a macroscopic manifold with a volume ${\cal{V}}$ located close to the photon sphere rim, the Hamiltonian unperturbed by  e-m field commutes with momentum operator and single-particle states $\vert 1(2)\rangle = \frac{1}{(2\pi\hbar)^{3/2}} e^{i (\mathbf{p}_{1(2)}\cdot \mathbf{r}-E_{p_{1(2)}}t)/\hbar}$ can represent states in the isotropic Fermi sphere if $p_{1(2)}\leq p_F$. 
The matrix element in (\ref{zrf}) is thus of Fourier component form, 
\begin{equation}
	\label{zrf1}
	\langle \mathbf{p}_1 \vert V(\mathbf{r})\vert \mathbf{p_2}\rangle =\mp \delta(\mathbf{p}_1-\mathbf{p}_2-\mathbf{q})
	\frac{ec^2\mathbf{A_0}\cdot \mathbf{p}_2}{\sqrt{p_2^2c^2+m^2c^4}},
\end{equation}
which gives for (\ref{zrf}),
\begin{equation}
	\label{zrf3}
	w_{1,2}=\frac{{\cal{V}}}{(2 \pi)^2\hbar}e^2c^2A_0^2cos^2\theta f(p_2)\delta(\mathbf{p}_1-\mathbf{p}_2-\mathbf{q})\delta(E_{p_1}-E_{p_2}-\alpha cq),
\end{equation}
where $f(p_2)=\frac{p_2^2c^2}{p_2^2c^2+m_{e(p)}^2c^4}<1$ and for $\delta^2(\mathbf{p})= \delta(\mathbf{p}=0)\delta(\mathbf{p})=\frac{\cal{V}}{(2\pi)^3}\delta(\mathbf{p})$ with ${\cal{V}}\rightarrow \infty$ (because $\delta(\mathbf{p}=0)=\frac{1}{(2\pi)^3}\int e^{i\mathbf{p}\cdot \mathbf{r}}d^3r=\frac{\cal{V}}{(2\pi)^3}$),
$E_{p_{1,2}}=\sqrt{p^2_{1(2)}c^2+m_{e(p)}^2c^4}-m_{e(p)}c^2$, $cos\Theta=\mathbf{A}_0\cdot \mathbf{p}_2/A_0p_2$. Because of $\alpha\simeq 0.57$, both Dirac deltas in (\ref{zrf3}) can be simultaneously fulfilled and transitions are allowed independently of $\mathbf{p}_1$. The probability $w_{1,2 }$ goes to zero for $\mathbf{p}_2\rightarrow 0$, but for $p_2^2 c^2\gg m_{e(p)}^2c^4$ the function $f(p_2)$ practically does not depend on $p_2$ as in this limit $f(p_2)\simeq 1$. For electrons (protons) this inequality is satisfied for $p_2>5 \times 10^{-22}$ ($5 \times 10^{-19}$) kg m/s. Higher energies dominate as the density of states in isotropic Fermi sphere grows as $p^2$.

Integration of Eq. (\ref{zrf3}) over all occupied states in the Fermi sphere allows the estimation of time-span for a Fermi sphere collapse. The integration over $\mathbf{p}_1$ and $\mathbf{p}_2$ gives from (\ref{zrf3}),
\begin{equation}
	\label{fermirule}
	\begin{array}{l}
		w= \int_{p_F}d^3p_1\int_{p_F}d^3p_2 w_{1,2}= (N+1)\gamma,\\
		\gamma = \frac{y^3 m^5 c^9 e^2}{12 \pi^2 \varepsilon_0 \hbar^3} \int_{-1}^1 dz\int_0^{p_F/mc}dx
		\frac{x^2}{x^2+1}\delta(\sqrt{x^2+y^2+2 x y z+1}-\sqrt{x^2+1}-0.57 y),\\
	\end{array}
\end{equation}
where $x=p_2/mc$, $y=q/mc=\hbar \omega/mc^2$, $\varepsilon_0$ is the dielectric constant and $N=\varepsilon_0 \frac{{\cal{V}}E_0^2}{2 \hbar\omega}$ is the number of photons $\hbar \omega$ in the volume ${\cal{V}}$, $E_0=V_0/\omega$ and the density of energy of e-m field is $\varepsilon_0 E_0^2 /2$. Nonzero value for $N=0$ reveals the spontaneous emission (as in optical transitions the only nonzero matrix element of the creation operator for photons is out of diagonal and equal to $\sqrt{N+1}$ for the quantized vector potential in (\ref{zaburzenie})). The number of photons satisfies the dynamic equation $dN=(N+1)\gamma dt$, hence, $ln N_m=\gamma\Delta t$, and $\Delta t$, the time of the Fermi sphere decay, equals to $ \frac{ln{N_m}}{\gamma}$ for particular $\hbar\omega= y mc^2$ and $N_m$ is of order of the total number of particles in the system. (Technical note: in (\ref{fermirule}) the integrals over $\mathbf{p}_1$ and $\mathbf{p}_2$ should be accompanied with the density of states, $\left(\frac{{\cal{V}}}{(2\pi \hbar)^3}\right)^2$, but the states $\vert 1(2)\rangle$ are normalised to the Dirac delta and to restore their probability sense, the matrix element (\ref{zrf1}) must be divided by $\frac{{\cal{V}}}{(2\pi \hbar)^3}$ to change the normalisation to volume ${\cal{V}}$; since the matrix element (\ref{zrf1}) enters into (\ref{zrf}) in the square, then the state density coefficients in (\ref{fermirule}) cancel.)

For a neutron star merger with $2.3$ Sun masses, one can estimate the time-span of the  decay of the Fermi sphere of electrons (assuming in (\ref{fermirule}) $m=9.1 \times 10^{-31}$ kg) and protons (ca. 2000 times more massive) for particular energy of emitted photons -- as shown in Table \ref{tab-kw2}, which agrees with the duration of short gamma-ray bursts. In this assessment we did not take into account the time of the decay of neutrons into electrons and protons, which additionally shifts the onset of the burst, but not its duration. Protons give the shorter burst, but their Fermi sphere total energy is only $\sim 0.7$ of the Fermi sphere energy of electrons (at the same Fermi momentum of both). Thus, the burst has high intensity during the first milliseconds in the higher photon energy range ($\hbar \omega >1$ MeV). The afterglow with lowering frequencies lasts longer the lower the photon energy is (for $\hbar\omega <0.01$ MeV longer than ca. $800$ s mostly due to electron transitions), which also remains in agreement with observations of such bursts.

\begin{table}
	\centering
	\begin{tabular}{c|c|c|c|c|c}
		\hline 
		& $0.01$ MeV & $0.1$ MeV & $1$ MeV & $10$ MeV& $100$ MeV\\
		\hline
		el&780 s&7.8 s& $7.7 \times 10^{-2}$ s& $7.5 \times 10^{-4}$ s& $6.4 \times 10^{-6}$ s\\
		\hline
		pr&$1.7 \times 10^{-3}$ s& $1.7 \times 10^{-5}$ s&$1.68 \times 10^{-7}$ s&$1.63 \times 10^{-9}$ s&$1.27 \times 10^{-11}$ s\\
		\hline
	\end{tabular}
	\caption{The Fermi golden rule estimation of the time of complete decay of the Fermi sphere for neutron star merger with the mass $2.3$ M$_{\odot}$, as in Table \ref{tab-kw1}, for different photon energies and for electron (el) and proton (pr) contributions.}
	\label{tab-kw2}
\end{table}

\section{Discussion and conclusion}

The disappearance of quantum statistics at passing the photon sphere rim in multi-particle systems approaching the event horizon of a black hole \cite{pra2023,jhep2022} explains the release of short giant gamma-ray bursts at collapses of neutron star mergers exceeding the TOV limit. Both the energy and timing of these bursts (of order of $10^{47}$ J and millisecond span of its high energy photon part emission) agree with the energy accumulated in the Fermi sphere of neutrons and with the timing of its decay due to local waiving off of Pauli exclusion principle. These seem to be not accidental, the more that the spectrum of radiation associated with the Fermi sphere decay is also congruent with the observed characteristics of short gamma-ray burst, both in the main high energy part within a few milliseconds and of lower energy afterglow in duration of several dozen of minutes. Moreover, no other mechanism is known to explain the conversion ratio up to 30 \% of mass (ca. one half of Sun mass at the collapse of 2.3 Sun masses at TOV limit) into radiation in so short time period.

The same quantum mechanism probably contributes  to the luminosity of largest quasars, which cannot be explained within conventional hydrodynamical models of accretion discs via their thermal radiation supplemented with an inverse comptomisation effect. The latter fits to the luminosity of small quasars with the central black hole of ca. $15$ Sun masses \cite{shapiro} and needs the temperature of an inner part of the accretion disc (terminated at ca. $6r_s$ from the singularity) on the level of $10^9$ K for electrons (and of $10^{11}$ K for ions) to elucidate the comptomisation of softer thermal photons. In the case of supermassive black holes of order of $10^9$ solar masses, a similar effect is not realistic, because the inner part of the accretion disk is too vast for such huge black holes and cannot be as hot as in the case of a microquasar. In the case of superluminous quasars the conventional hydrodynamic model of accretion disc (supported by the comptomisation effect) is thus insufficient \cite{supereddington,shapiro,6p,24p} to explain the luminosity of order of $10^{40}$ W, with high energy photon contribution (up to GeV). The decay of Fermi spheres of electrons and protons in plasma passing the photon sphere rim in the distance of $1.5r_s$ from the central singularity resolves these problems out of reach for conventional hydrodynamic models, and supplements the total radiation of superluminous quasars just in the gamma part of the spectrum and with sufficient radiation intensity. The quantum statistics effect has up to 30 \% efficiency of mass to energy conversion and does not conflict with additional radiation from the more distant parts of the vast accretion disc described so-far upon conventional classical approach \cite{sunaev,novikov,shapiro,farahzest,6p,24p}.

So high efficiency of mass to radiation energy conversion needs the uppermost density of plasma compressed by the supermassive black hole without any restrictions imposed on the matter supply to the disc, or  the conditions similar to those occurring in unstable neutron stars passing the TOV stability limit. In the case of not extremal compression of matter falling into black hole the effect is similar, though with much lower efficiency for mass to radiation conversion. This however, seems to fit to some observations of transients in active galactic nuclei (AGN) or of frequently observed tidal disrupt events (TDE) of much smaller intensity. The application of the Fermi sphere collapse at passing the photon sphere rim conveniently supplements the elucidation of AT 2020neh one year-lasting episode classified as TDE \cite{ggg50000} of a main sequence star (with $1.3$ Sun mass) by the central hole of a dwarf galaxy SDSSJ152120.07+140410.5 (with $z=0.062$ and mass similar to Great Magellan Cloud). The luminosity, maximal of order of $4 \times 10^{36}$ W, lasting shorter than a month, was concentrated in UV and optical range in agreement with conventional models of TDE, though with some contribution in X-ray range of $4.5 \times 10^{34}$ W exceeding these models. This can be, however, explained by inclusion of energy released due to the Fermi sphere collapse at passing the photon sphere rim by the star debris at TDE with the compression of fermions adjusted to the greatest observed energy of photons.

 The recently observed for AGN 1ES 1927+654 \cite{ga1flare} 100-fold increase of its luminosity within a few months period is probably associated with accidental increase of the matter consumption rate during this period. The matter influx was not extremal and photons emitted due to the Fermi sphere collapse did not reach over-MeV energy and were not able to produce electron-positron pairs in the ergosphere of this spinning black hole. However, the massive isotropic flux of lower energy photons caused by the Fermi sphere collapse of electrons and protons pushed electron-positron pairs created in the ergosphere according to Blandford-Znajek electromagnetic mechanism \cite{znajek} towards the event horizon, lowering in this way their evaporation to jets across ergosphere nodes and reducing positron and electron population in jets. This can explain \cite{jcap} the temporal change in the radiation spectrum during brightening episode AGN 1ES 1927+654 consisting in 100-fold increase of the optical luminosity and simultaneous lowering of X-ray radiation (the latter probably due to the reduction of the amount of electrons and positrons in jets, the source of X-ray radiation here), without a need to speculate on re-magnetization of the AGN and quenching its jets by oppositely magnetised gas cloud during this episode \cite{ga1flare}.

Finally, the described in the present paper contribution to the problem of information paradox and related concept of the firewall of a black hole seems to be the theoretical support for the model of quantum statistics disappearance in close vicinity of the event horizon. The clarification of the information paradox is important especially in view of the related breaking of the Einstein equivalence principle, unitarity, or existing quantum field theory \cite{hawking,polczinski,maldac}. The described quantum property of black holes rooted in topology of trajectories close to the general relativistic gravitation singularity removes the paradox and paves a new way to avoid related problems with hypothetical firewall \cite{polczinski}. Because of the absence of quantum statistics beneath the photon sphere rim of any black hole, the quantum field theory in this region must be verified -- it cannot be neither of fermionic nor of bosonic type. This needs a theoretical development exceeding the scope of the present paper, but it poses a question to what extent such a statistics-free field approach would help in avoidance of problems with quantisation of gravity.

\section{Methods }
\subsection{Trajectory homotopy close to the photon sphere rim}
\label{aaa}
To notice the change of trajectory homotopy class at the photon sphere rim let us consider the availability  of closed loops for trajectories in various regions of the upper vicinity of the event horizon. This resolves itself to the consideration of geodesics in the metric (\ref{metryka1}). The geodesics can be defined as the solution of the Hamilton-Jacobi equation,
\begin{equation}
	\label{geodesics}
	g^{ik}\frac{\partial S}{\partial x^i}\frac{\partial S}{\partial x^k}-m^2c^2=0,
\end{equation}
with $g^{ik}$ metric tensor components corresponding to metric (\ref{metryka1}) \cite{lanfielda}.
Eq. (\ref{geodesics}) attains for the Schwarzschild metric (\ref{metryka1}) the following form,
\begin{equation}
	\label{jacobi1}
	\begin{array}{l}
		\left(1-\frac{r_s}{r}\right)^{-1}\left( \frac{\partial S}{c\partial t}\right)^2\\
		-\left(1-\frac{r_s}{r}\right)\left(\frac{\partial S}{\partial r}\right)^2
		-\frac{1}{r^2}\left(\frac{\partial S}{\partial \phi}\right)^2-m^2c^2=0,\\
	\end{array}
\end{equation}
with the	function $S$ in the form,
\begin{equation}
	\label{jacobi}
	S=-{\cal{E}}_0t+{\cal{L}}\phi+S_r(r).
\end{equation}
In the above formula the quantities ${\cal{E}}_0$ and ${\cal{L}}$ are the particle energy and its angular momentum, respectively. ${\cal{E}}_0$ and ${\cal{L}}$ are constants of motion.
Eq. (\ref{geodesics}) can be also applied to define trajectories of photons assuming in (\ref{geodesics}) $m=0$.

If one substitutes Eq. (\ref{jacobi}) into Eq. (\ref{jacobi1}), then one can find $\frac{\partial S_r}{\partial r}$. By the integration of this formula one obtains, 
\begin{equation}
	\begin{array}{l}	
		S_r=\int dr\left[ \frac{{\cal{E}}_0^2}{c^2}\left(1-\frac{r_s}{r}\right)^{-2} \right.\\
		\left.	-\left(m^2c^2
		+\frac{{\cal{L}}^2}{r^2}\right)\left(1-\frac{r_s}{r}\right)^{-1}\right]^{1/2}.\\
	\end{array}
\end{equation}

Geodesics are thus defined by the condition $\frac{\partial S}{\partial{\cal{E}}_0}=const.$, which gives radial dependence of the trajectory $r=r(t)$, and by the condition $\frac{\partial S}{\partial{\cal{L}}}=const.$, determining the angular dependence $\phi=\phi(t)$ of the particle trajectory. 
The condition $\frac{\partial S}{\partial{\cal{E}}_0}=const.$ gives,
\begin{equation}
	\label{promien}
	ct=\frac{{\cal{E}}_0}{mc^2}\int \frac{dr}{(1-\frac{r_s}{r})\sqrt{\left(\frac{{\cal{E}}_0}{mc^2}\right)^2-\left(1+\frac{{\cal{L}}^2}{m^2c^2r^2}\right)\left(1-\frac{r_s}{r}\right)}}.
\end{equation}
The condition $\frac{\partial S}{\partial{\cal{L}}}=const.$ results in the relation,
\begin{equation}
	\label{phase}
	\phi=\int dr\frac{{\cal{L}}}{r^2}\left[\frac{{\cal{E}}_0^2}{c^2}-\left(m^2c^2+\frac{{\cal{L}}^2}{r^2}\right)
	\left(1-\frac{r_s}{r}\right)\right]^{-1/2}.
\end{equation}

Eq. (\ref{promien}) can be rewritten in a differential form,
\begin{equation}
	\label{differential}
	\frac{1}{1-r_s/r}\frac{dr}{cdt}=\frac{1}{{\cal{E}}_0}\left[{\cal{E}}_0^2-U^2(r)\right]^{1/2},
\end{equation}
with the effective potential,
\begin{equation}
	\label{potencjal}
	U(r)=mc^2\left[\left(1-\frac{r_s}{r}\right)\left(1+\frac{{\cal{L}}^2}{m^2c^2r^2}\right)\right]^{1/2},
\end{equation}
where ${\cal{E}}_0$ and ${\cal{L}}$
are energy and angular momentum of the particle, respectively.

The equation (\ref{differential}) allows for the definition of an accessible region for the motion via the following condition, ${\cal{E}}_0\geq U(r)	$.
Moreover, the condition 
${\cal{E}}_0= U(r)	$ defines circular orbits.
Limiting circular orbits can be thus found by the determination of extrema of $U(r)$. Maxima of $U(r)$ define unstable orbits, whereas minima stable ones (depending on parameters ${\cal{E}}_0$ and ${\cal{L}}$, which are integrals of the motion). The conditions $U(r)={\cal{E}}_0$ and $\frac{\partial U(r)}{\partial r}=0$ (for extreme) attain the explicit form,
\begin{equation}
	\label{branches}
	\begin{array}{l}
		{\cal{E}}_0={\cal{L}}c\sqrt\frac{2}{rr_s}\left(1-\frac{r_s}{r}\right),\\
		\frac{r}{r_s}=\frac{{\cal{L}}^2}{m^2c^2r_s^2}\left[1\pm \sqrt{1-\frac{3m^2c^2r_s^2}{{\cal{L}}^2}}\right],
	\end{array}
\end{equation}
where the sign $+$ in the second equation corresponds to stable orbits (minima of $U(r)$) and the sign $-$ to unstable ones (maxima of $U(r)$). Positions of stable and unstable circular orbits depend on energy ${\cal{E}}_0$ and angular momentum ${\cal{L}}$. 
This is illustrated in Fig. \ref{photosphere} -- the upper curve (red one in this figure) gives positions of stable circular orbits (with respect to angular momentum ${\cal{L}}$) and the lower curve (blue one) gives positions of unstable circular orbits (also with respect to ${\cal{L}}$). The related value of ${\cal{E}}_0$ is given by the first equation of the system (\ref{branches}).

The spiral trajectories described by equations (\ref{promien}) and (\ref{phase}) are short and depend on motion integers ${\cal{L}}$ and ${\cal{E}}_0$, the angular momentum and the energy of a particle, respectively. The effective potential of a black hole (\ref{potencjal}) defines the accessible movement region by the condition ${\cal{E}}_0\geq U(r)$. As we consider particles crossing the photon sphere rim and the maximum of $U(r)$ is located just around this sphere, one can estimate accessible ${\cal{E}}_0$ for particular ${\cal{L}}$. For such a pairs of ${\cal{L}}$ and ${\cal{E}}_0$, the phase shift of the spiral between the photon sphere rim and the event horizon, given by the integral (\ref{phase}) is confined to rather small values, what is visualised in Fig. \ref{trajektorie-bh}.

\subsection{Entropy}

The entropy is defined as the equilibrium state function of thermodynamic systems and is extended also onto  quantum states. The entropy of a pure quantum state is zero and for mixed states is non-zero. A mixed state can be described by the density matrix operator $\hat{\rho}= Tr'\vert \Psi\rangle \langle\Psi\vert$, where $\vert \Psi\rangle \langle \Psi\vert$ is the projection operator onto the pure state of the total system consisting of the subsystem and the surroundings and $Tr'$ goes here over the Hilbert space of the surroundings. The entropy for the subsystem is defined as $S=-Tr\hat{\rho} ln \hat{\rho}$ and $Tr$ goes here over the Hilbert space ${\cal{H}}_1$ of the subsystem (the whole system Hilbert space ${\cal{H}}$ is the tensor product ${\cal{H}}={\cal{H}}_1\otimes {\cal{H}}_2$, with ${\cal{H}}_2$ -- the Hilbert space of the surroundings). Interaction between subsystems usually leads to the quantum entanglement, i.e., the pure state of the total system is not a factorised product of pure states of components, but is an element from the tensor product space ${\cal{H}}$ which does not admit factorization. An example of such a situation is the subsystem in equilibrium described by the Gibbs canonical ensemble with $\hat{\rho}= \frac{e^{-\beta \hat{H_1}}}{Tre^{-\beta\hat{H_1}}}$ or grand-canonical ensemble $\hat{\rho}=\frac{e^{-\beta(\hat{H_1}-\mu\hat{N_1})}}{Tre^{-\beta(\hat{H_1}-\mu\hat{N_1})}}$, where $\beta=\frac{1}{k_BT}$ ($k_B$ is the Boltzmann constant, $T$ is the absolute temperature) and $\mu$ is the chemical potential -- they are the Lagrange factors to maximize the information entropy in the equilibrium state at $Tr\hat{\rho}\hat{H_1}=E= constant$ and $Tr \hat{\rho}\hat{N_1}=N=constant$ (${\hat{N_1}}$ is the operator of particle number in the subsystem). The two latter conditions define conservation of the energy and particle number in mean value, respectively.
The canonical or grand-canonical ensembles describe, according to Gibbs theorem, small subsystems of a larger one, which is in a pure but entangled state.

For another example let us take the system of $N$ identical indistinguishable particles satisfying the quantum statistics of bosonic or fermionic type. In both these cases the multi-particle wave function of the total $N$-particle system is entangled in the Hilbert space ${\cal{H}}=\prod_{i=1}^N\otimes {\cal{H}}_i$, where ${\cal{H}}_i$ is the individual Hilbert space for $i$-th particle. Even if particles do not interact, they are maximally entangled due to the requirement of symmetry or antisymmetry of the wave function at particle interchanges, for identical indistinguishable bosons or fermions, respectively. Entropy (the von Neumann entropy) of each particle is non-zero as for its mixed state. The organisation of a matter due to interaction (reducing the entropy and enhancing the information encoded) dominates in fermionic structures of elementary particles. The energy stored in such systems is conditioned by  Pauli exclusion principle for fermions. Also the core repulsion of hadrons is originated by their fermionic quark internal contents.
Structure and energy of atoms and molecules is conditioned by Pauli exclusion principle, as well as the energy of Fermi spheres of high density systems of fermions. When quantum statistics is degraded, then these energies convert into e-m radiation along the Fermi golden rule for quantum transition of charged particles \cite{landau1972a}. The matter loses its information and entropy.

Super-luminous quasars with the luminosity of $\sim 10^{40}$ W or some kind of giant gamma-ray bursts with energy of $\sim 10^{47}$ J supposed to accompany collapses of neutron star mergers, suggest an enormously efficient processes with up to $30$ \% mass to radiation energy conversion ratio. The nuclear fusion in stars is of only $0.7$ \% mass to energy conversion efficiency and the perfect $100$ \% efficiency of annihilation does not contribute here as it would need the unrealistic supply of large amount of antimatter. The luminosity of quasars has been explained via the accretion of matter \cite{salpet,zeld,ksiazka} from surroundings of a central black hole on the cost of its gravitational energy eventually converted into thermal radiation from the hot accretion disc \cite{sunaev,novikov,merloni} and additionally strengthened by inverse Compton, bremsstrahlung and synchrotron effects. The realistic assessment of related luminosity has been supplied \cite{shapiro} but limited to small black holes (in \cite{shapiro} for micro-quasar Cignus X-1 with  15 Sun mass black hole). This model is hard to be scaled to super-luminous quasars powered by black holes with the mass of order of $10^9$ Sun masses \cite{supereddington,6p,farahzest,24p}. In the case of giant gamma-ray bursts of neutron star mergers exceeding the TOV limit \cite{tolman,volkoff}, none of the known conventional mechanisms is able to explain the release of short radiation impulse with energy $\sim 10^{47}$ J equivalent to ca. $0.5$ Sun mass.

We propose a supplementary mechanism of the gravitation energy conversion into radiation via a quantum channel, which is active in close vicinity of the event horizon of a black hole, and which, at certain conditions, can achieve the efficiency up to $30$ \% mass to radiation energy conversion ratio. This channel can contribute to conventional models of accretion disc \cite{sunaev,novikov,shapiro} by large non-thermal radiation from the vicinity of the event horizon inaccessible to classical hydrodynamic models of accretion and matching to observations of super-luminous quasars. The same mechanism can explain the giant gamma-ray bursts suspected to be released at neutron star unstable merger collapses (cf. \ref{F}). The proposed mechanism is rooted in topology of the curved spacetime close to the gravitational singularity, is an universal property of each black hole and manifests itself at episodes of the matter consumption by black holes. The decay of quantum statistics and the collapse of the matter structured due to Pauli exclusion principle releases the accumulated energy and the entropy/information in an unitary manner according to the Fermi golden rule. The energy and entropy is taken away by emitted radiation. The remaining matter continues to fall to the event horizon (and crosses it in a finite proper time) but is devoid of quantum statistics and has no way of exchanging the positions of indistinguishable particles, whether they were previously fermions or bosons. This state is similar to an ideal Nernst crystal with zero entropy and zero information as in the third law of thermodynamics.

\subsection{Pauli theorem on connection between statistics and spin}
\label{C}
Quantum statistics is a property of collective multi-particle systems and is assigned to particles by 1DUR of the braid group of the system. It cannot be assigned to a single particle because the braid group requires at least two particles to be exchanged \cite{mermin1979}. The same classical particles can acquire various quantum statistics depending on the manifold in which they are placed and on the choice of 1DUR (usually in multiple numbers for the same braid group). The selected 1DUR governs over the multi-particle wave function of the system, which must transform as the 1DUR of the braid when the arguments of this function are exchanged in the manner prescribed by this particular braid -- the element of the braid group. For 3D manifolds $M$ the braids are elements of the permutation group, but for 2D -- not. In the latter case the rich structure of braids is the topological reason of exotic manifestation of fractional quantum Hall effect \cite{annals2021}.

The selection of 1DUR for a specific particle system undergoes according to the Pauli theorem on spin and statistics \cite{pauli}, which asserts that particles with half spin must be fermions, whereas those with integer spin -- bosons. 
The reasoning standing originally behind this theorem is related with the Dirac electrodynamics for $\frac{1}{2}$ spin particles, which admitted the Hamiltonian formulation and for positive definition of the kinetic energy of free particles and antiparticles the corresponding to them quantum fields must anti-commute like for fermions. The generalisation of such a reasoning for interacting particles has not been achieved \cite{duck1,duck2}, which suggests that the proof of the Pauli theorem should be rather completed in topological terms in view of by homotopy group quantum statistics definition \cite{leinaas1977,sud,lwitt-1}. 
Within this approach, Pauli theorem follows from the coincidence of unitary irreducible representations of the rotation group, which define quantization of spin or angular momentum \cite{rumerfet1}, with unitary representations of braid groups nominating quantum statistics \cite{mermin1979,lwitt-1}. The agreement between unitary representations of both groups arises due to the overlap of some elements of the braid group and the rotation group. The representations which are uniform on group generators must thus agree for whole groups. For 3D manifolds, the rotation group $O(3)$ has the covering group $SU(2)$ and the irreducible unitary representations of $SU(2)$ fall into two classes assigning integer and half-integer angular momenta. These two classes agree with only two possible scalar unitary representations of the permutation group $S_N$, which is the braid group for 3D manifolds. The representations of both groups coincide as they have some common elements.
The half spin representation of the rotation group must agree with the odd representation $\sigma_i\rightarrow e^{i\pi}=-1$ of the braid group for 3D manifold (fermions) ($\sigma_i$ are generators of the permutation group which is a braid group in 3D), whereas the integer angular momentum representation of the rotation group must agree with even representation $\sigma_i\rightarrow e^{i0}=1$ of the permutation group (bosons).
The topological approach allows for the extension of Pauli theorem onto 2D manifolds with anyons (which, in general, are neither fermions nor bosons \cite{wilczek}). For 2D manifolds the rotation group $O(2)$ is Abelian and isomorphic with $U(1)$ group possessing just the same continuous unitary representations $e^{i\alpha}$, $\alpha\in[0,2\pi)$, as the Artin group, which is the braid group for $M=R^2$. Thus, in two dimensional space Pauli theorem also holds for not quantized spin assigned by $s=\frac{\alpha}{2\pi}$ and similarly continuously varying anyon statistics defined by $e^{i\alpha}$ numbered by $\alpha\in[0,2\pi)$.

Hence, it is clear that quantum statistics and spin, though coincide via the agreement between unitary representations of rotation and braid groups, are in fact independent. It is possible a situation when the spin is still defined but the statistics  not, as in the case of the absence of a nontrivial braid group for the simply-connected configuration space $F_N$, as it occurs beneath the photon sphere rim of a black hole.

\subsection{Existence of braids and availability of trajectories for Feynman path integration}

\label{ggg1}

Rigorous mathematical formulation of quantum statistics involves Feynman path integration \cite{feynman1,feynman1964} generalised for multiparticle systems of identical indistinguishable particles \cite{lwitt,chaichian1,chaichian2}.

The path integral for a single particle originally proposed by Feynman \cite{feynman1964} has the form,
\begin{equation}
	\label{gwn500}
	I(z_1,t_1;z_2,t_2)=\int d\lambda e^{i \int_{t_1}^{t_2}{\cal{L}}[\lambda(z_1,t_1;z_2,t_2)]/\hbar},
\end{equation}
where the summation (integration with a measure $d\lambda$) over trajectories concerns all accessible classical trajectories $\lambda$ linking the initial point $z_1$ in the configuration space of this particle at time $t_1$ and the final point $z_2$ at time instant $t_2$. Lagrangian ${\cal{L}}=T-V$ integrated over time gives the action $S[\lambda]$ -- a functional over the domain of trajectories. 
This functional is minimal (extremal) for classical trajectory defined by the Euler-Lagrange equation, $\frac{d}{dt}\left(\frac{\partial \cal{L}}{\partial \dot z}\right)-\frac{\partial {\cal{L}} }{\partial z }=0$, according to the least action principle. 
The family of trajectories contributing to Feynman path integral includes also arbitrary not extremal paths, but those which are classically accessible for the potential and for the topology of the configuration space. Quantum Feynman path integral is analogous to the Wiener classical path integral \cite{wienera} applied to the Brownian motion \cite{chaichian1}. In distinction to the well defined Wiener measure for summation of probabilities \cite{wienera,chaichian1}, the summation of complex probability amplitudes in quantum Feynman integral \cite{feynman1964} precludes a proper measure definition \cite{chaichian1}, which admits only a more heuristic approach consisting in the discretization method for explicit summation over trajectories \cite{feynman1964,chaichian1}. This method shows, however, which trajectories contribute to the sum -- arbitrary continuous piecewise paths of segments being minimal (extremal) for boundary conditions on discrete consecutive internal points (approximated by straight line sectors, without any loss of generality at limiting infinitesimal time-step of discretization). This needs, however, the existence of a real extreme trajectory for each segment. With this restriction imposed onto the discretization procedure, the integration over all internal points is performed. For quadratic Langrangians these integrals are of Gaussian type, which attain the analytic form only for infinite limits of integration. However, if the piecewise trajectory cannot be constructed in some coordination space region (due to features of the potential $V(z)$ or topological restrictions imposed) the integral limits cannot always be infinite. This problem has been noticed by Pauli \cite{paulib,chaichian1} and also earlier for Wiener path integrals for Brownian motion with inaccessible regions \cite{chaichian1}.

To exemplify the situation when classical trajectories are inaccessible, one can consider quantum tunnelling across the classically inaccessible region. This tunnelling is the same in Schr\"odinger quantum mechanics formulation or by Feynman path integral quantisation. The tunnelling occurs for a finite height or non-vertical (though infinite) barrier. In the case of an infinitely high barrier none classical trajectories beyond the barrier exist and the propagator across the barrier is zero \cite{gwn102}, which is equivalent to the fact that quantum particles do not tunnel through infinitely high potential barriers and cannot penetrate such infinite vertical potential walls. This simple example shows that not all trajectories contribute to the Feynman path integrals, but only those which are classically accessible (even piecewise) for a given potential and topological restrictions. For e.g., 1D oscillator potential $x^2$ is the parabola of infinite height, but the potential is not vertical and it does not restrict the possibility of arbitrary piecewise trajectory construction (as the parabola is infinitely wide). However for a rectangular 1D infinite well or potentials ranged by vertical asymptotes, trajectories linking both separated sides are not admitted and quantum propagator is zero, with exception for an infinite potential of Dirac delta type, which despite its infinity allows a nonzero quantum transition across it. The Dirac delta is nonzero only in a single point and this point can be selected as fixed one of the discretization points, which does not conflict with the existence of trajectories on the left and on the right, and using the property for multiplication for path integral \cite{feynman1964} ch. 2-5 (with integration over the intermediate time), one gets nonzero probability transmission from the left to the right. For a finite width infinite barrier such  argumentation cannot be applied and the propagator between separated sides is zero.

The more complicated situations with inaccessible trajectories occur in collective multiparticle systems of identical particles. 
The path integral for $N$ indistinguishable identical particles takes the form \cite{lwitt,chaichian1,chaichian2,pra2023}, 
\begin{equation}
	\label{gwn501}
	I(Z_1,t_1;Z_2,t_2)=\sum_l e^{i\alpha_l}\int d\lambda_l e^{i S[\lambda_l(Z_1,t_1;Z_2,t_2)]/\hbar},
\end{equation}
where 
points $Z_1=(z_1^1,\dots, z_N^1) $ and $Z_2=(z_1^2,\dots, z_N^2)$ are two {\it different} points in multidimensional configuration space $F_N$ of $N$ indistinguishable particles, which define the start and final points for the propagator $I(Z_1,t_1;Z_2,t_2)$ at time instants $t_1$ and $t_2$, respectively. This propagator is the matrix element of the quantum evolution operator for a whole system in the position representation (between positions $Z_1$ and $Z_2$). The square of modulus of the complex propagator gives the probability of quantum transition between initial and final points at $t_1$ and $t_2$, respectively. 
The discrete index $l$ enumerates braids in the full braid group and $e^{i \alpha_l}$, $\alpha_l\in[0,2\pi)$ is the scalar unitary representation of $l$-th braid. The attachment of a braid loop to an arbitrary open trajectory between $Z_1$ and $Z_2$ reflects the possibility of particle numbering changes on the way in an arbitrary intermediate trajectory point. Braids are non homotopic, thus the measure in the path space can be only defined separately on each disjoined sector of the path space numbered by $l$ -- such a family of measures for path integration is denoted by $d\lambda_l$. The contributions to path integral from all sectors of the path domain numbered by braid group elements must be added with arbitrary unitary phase factor (unitarity is required to be consistent with quantum mechanics causality). The unitary factors are scalar unitary representations of the full braid group \cite{lwitt}.

The braids attached to an open multi-strand trajectory also must be built of classically accessible trajectories like all paths for Feynman integral over trajectories. The representation $e^{i\alpha_l}$ defines quantum statistics of particles. For multicyclic groups like braid groups, the 1DUR is sufficient to be defined on the group generators. This representation on generators does not depend on particular index $l$ of the generator -- in 3D braids are permutations and their unitary scalar representations are uniform over generators, $e^{i\pi}$ for fermions and $e^{i0}$ for bosons. In 2D anyons are assigned to also uniform representations $e^{i\alpha}$ independent of generator index. The latter property is visible from the upper relation (\ref{relacje}), because scalar representations 1DUR commute and from (\ref{relacje}) 1DUR($\sigma_i$)=1DUR($\sigma_{i+1}$), which gives the independence of the generator index. The same classical particles may have different quantum counterparts in dependence of the choice of an admissible unitary representation of the related braid group.

In the situation when classical trajectories for particle exchanges do not exist, it is impossible to define braids except  for a trivial one $e$. In such a case the summation over $l$ in (\ref{gwn501}) disappears and none quantum statistics can be assigned. Some external topological factors can also modify braids, which causes changes in their scalar unitary representations and thus in quantum statistics. This has the experimentally observed consequences in 2D electron systems exposed to a perpendicular magnetic field, which due to a classical cyclotron effect imposes restrictions on trajectories. In the case of a strong magnetic field, when the size of  cyclotron orbits is smaller than electron separation (due to the repulsion of electrons they form at $T=0$ K a rigid classical Wigner lattice on a uniform positive jellium, defining electron separation), which precludes electron exchanges, unless multi-loop cyclotron orbits are admitted. The latter have, however, different scalar unitary representations of related braids, which leads to the exotic phenomena of fractional quantum Hall effect \cite{tsui1982,laughlin2,annals2021,pra2023}.

In the case of trajectory homotopy class occurring at the vicinity of the event horizon of a black hole, for multi-particle systems passing the photon sphere rim, braids cannot be constructed from pieces of spirals given by Eqs. (\ref{promien}) and (\ref{phase}), except for trivial braid $e$. The multi-particle configuration space is simply-connected there and none quantum statistics can be assigned to particles beneath the photon sphere rim.

\subsection{Assessment of the efficiency of the collapse of Fermi spheres of electrons and protons in the accretion disc of a quasar including general-relativistic corrections}
\label{E}
The matter falling onto the black holes of super-luminous quasars must convert up to ca. 30 \% of their mass into radiation to explain their observable luminosity and simultaneously the rate of the increase of central black hole mass over a long time period of their activity to be consistent with observed masses of supermassive black holes in galaxies. As the giant black holes in closer galaxies are of size at most of the order of billions  masses of the Sun, the estimation of the mass consumption rate for quasars with luminosity of order of $10^{40}$ W is typically ca. $10$ Sun mass per year (i.e., of order of $0.1$ Earth mass per second), in extreme case of $1000$ Sun mass per year ($10$ Earth mass per second).

Central black holes in quasars vary between $ 10^6 - 10^9$ of solar masses, as have been measured using a reverberation mapping. Several dozen nearby large galaxies, including our own Milky Way, that do not have active centres and do not show any activity similar to a quasar, are confirmed to contain similar supermassive black holes in their centres. Thus it is now thought that all large galaxies have giant black holes of this kind, but only a small fraction have sufficient matter in the right kind of orbit at their centre to become active and power the radiation in such a way as to be seen as quasars.

For the concreteness of the estimation let us assume that the central black hole in quasar consumes 5.6 $\text{M}_{\odot}$ per year, i.e., ca. $0.06$ Earth mass per second. Let us assume the stable uniform in time process of matter accretion. The transport of matter across the disk is steady, thus we can perform calculation e.g., per a single second. Using Eqs (\ref{fm}) and (\ref{en}) one can assess the energy stored in the Fermi spheres for electrons and protons, if all the electrons and protons from the gas mass equalled to $0.06 $ Earth mass, are compressed to the spatial volume $V$ per second. The local Fermi momentum,
\begin{equation}
	\label{fmm}
	p_F(r)=\hbar(3\pi^2 \rho(r))^{1/3}=\hbar\left(3\pi^2 \frac{n}{V(r)}\right)^{1/3},
\end{equation}
where $r$ is the distance from the centre. $p_F(r)$ is constant in time and grows across the disk with increasing local concentration $\rho(r)=\frac{dn}{dV}=\frac{n}{V(r)}$, the same for electrons and protons. The latter equality holds for steady accretion and $n$ is the total number of electrons (or protons) per second, compressed in total to the volume $V(r)$ at the distance $r$ from the origin with central gravitational singularity. This means that portions $dn$ of electrons and protons in infinitely small consecutive periods $dt$ incoming in radial direction towards the central singularity compressed to $dV$ at radius $r$ add up in due of a single second time period to the total constant flow of mass (in the example, of $0.06 \times M_Z$ kg/s, the Earth mass $M_Z=5.97 \times 10^{24}$ kg) and as the whole is compressed locally at $r$ to the volume $V(r)$. The locally accumulated energy in the Fermi spheres of electrons and protons grows with lowering $r$ due to the increase of the compression caused by the gravitational field. This energy is proportional to $V(r)$ and, moreover, depends on $V(r)$ via the local Fermi momentum (\ref{fmm}) and in accordance with Eq. (\ref{en}) can be written as,
\begin{equation}
	\label{enn}
	\begin{array}{l}
		E(r)=E_e(r)+E_p(r),\\
		E_{el}(r)=\frac{V(r)}{2 \pi^2 \hbar^3}\int_o^{p_F(r)}dp p^2 \left( \sqrt{p^2 c^2 +m_e^2 c^4}-m_e c^2\right),\\ 
		E_{pr}p(r)=\frac{V(r)}{2 \pi^2 \hbar^3}\int_o^{p_F(r)}dp p^2 \left(\sqrt{p^2 c^2 +m_p^2 c^4}-m_p c^2\right),\\
	\end{array}
\end{equation}
where the energy $E_{el(pr)}$ refers to electrons (protons).

At the critical radius $r^*$ close to Schwarzschild horizon ($r^*=1.5r_s$), the decay of quantum statistics takes place due to the topological reason and both Fermi spheres of electrons and protons collapse. The amount of energy given by Eq. (\ref{enn}) per one second can be thus released in the vicinity of the Schwarzschild horizon due to collapse of Fermi spheres. This released energy per second can contribute in part to the observed luminosity of quasar ($10^{40}$ W). This undergoes by portions $dn$ of particle flow incoming to $r^*$ region in infinite small time periods $dt$, adding up in total to $0.06 \times M_Z$ per second. The value of the energy released depends on local Fermi momentum and attains $10^{40}$ J at sufficiently high level of compression, i.e., at sufficiently small $V(r^*)$ determined from the self-consistent system of Eqs (\ref{fmm}) and (\ref{enn}) if one assumes $E(r^*)=10^{40}$ J.

To the initial mass of a gas (assuming to be composed of hydrogen H) contribute mostly protons (ca. 2000 times more massive than electrons), thus the total number of electrons, the same as the number of protons, equals to, $n\simeq 0.06 M_Z /m_p\simeq 2.14 \times 10^{50}$ per second. Simultaneously solving Eqs (\ref{fmm}) and (\ref{enn}), assuming $n=2.14 \times 10^{50}$ in volume $V(r^*)$ and released energy $E(r^*)=10^{40}$ J, we find the volume of compressed plasma $V(r^*)=0.5 \times 10^5$ m$^3$ and electron or proton Fermi sphere radius $p_F(r^*)=5.4 \times 10^{-19}$ kg m/s. Electrons and protons (their amount per second) are compressed to the same volume $V(r^*)$ (due to neutrality of plasma), hence, their concentration at $r^*$, $\rho(r^*)=4.3 \times 10^{45}$ 1/m$^3$. The mass density at $r^*$ (including the mass equivalent to the energy stored up in Fermi spheres of electrons and protons) is thus $\xi(r^*)=\frac{0.06 M_Z}{V(r^*)}+\frac{E(r^*)}{c^2 V(r^*)}\simeq 9 \times 10^{18}$ kg/m$^3$, similar to mass density in neutron stars at TOV limit (being of order of hadron density in atom nuclei). This is the uppermost mass density at the critical $r^*$, which evidences the self-consistency of the model. This limit regulates the matter consumption by a black hole, when the supply of the matter to an accretion disc is unlimited in the black hole surroundings. The released energy of $E(r^*)=10^{40}$ J is equivalent to 30 \% of the falling mass of $0.06$ Earth mass (per second). It means that the compressed plasma with degenerate Fermi liquid of electrons (and also of protons) is at $r=r^*$ by 30 \% more massive than initial remote diluted gas. This increase of mass is caused by the gravitational field of the central black hole, which compresses both systems of fermions accumulating the energy in their Fermi spheres.

The ratio of total Fermi sphere energies of electrons and protons is $\frac{E_{el}(r^*)}{E_{pr}(r^*)}\simeq 1.4$. The Fermi energy of electrons with Fermi momentum $p_F(r^*)=5.48 \times 10^{-19}$ kg m/s equals to $\varepsilon_F= 1 $ GeV (it is the uppermost possible energy of emitted photons), which in thermal scale (in units of $k_B=1$) is of order of $10^{13}$ K -- this makes the electron liquid quantumly degenerated at lower temperatures (quasars are not source of thermal gamma radiation, thus their actual temperatures are much lower). The Fermi energy of protons with Fermi momentum $p_F(r^*)=5.48 \times 10^{-19}$ kg m/s equals to $\varepsilon_F= 0.4 $ GeV (it is the uppermost possible energy of emitted photons by jumping of protons), which in thermal scale (in units of $k_B=1$) is of order of $4 \times 10^{12}$ K -- thus for the temperature of plasma of order of $10^{6-9}$ K (more realistic is $10^6$) protons also form the degenerated Fermi liquid.

The release of energy due to the collapse of the Fermi sphere of charged particles undergoes according to the Fermi golden rule scheme for quantum transitions \cite{landau1972a}, when such transitions are admitted by the local revoking of Pauli exclusion principle. Charged carries (electrons and protons) couple to electromagnetic field and the matrix element of this coupling between the individual particle state in the Fermi sphere and its ground state is the kernel of the Fermi golden rule formula for transition probability per time unit for this particle.
This interaction depends also on electromagnetic field strength, thus the increasing number of excited photons strengthens the coupling in the similar manner as at stimulated emission (known from e.g., laser action) and accelerates quantum transition of the Fermi sphere collapse (Method \ref{F}).

Note that the above estimation of the energy accumulated in Fermi spheres at critical distance from the gravitational singularity has been done in conventional rigid coordinates, time and space-spherical coordinates $(t,r,\theta,\phi)$ of the remote observer. Schwarzschild metric (\ref{metryka1}), though written in the rigid and stationary coordinates, describes the folded spacetime.
Even if in the Eq. (\ref{fmm}) one replaces $V(r)$ by the proper volume at the distance $r$ from the central singularity, then according to the Schwarzschild metric (\ref{metryka1}) one obtains for the elementary proper volume the formula,
\begin{equation}
	\label{correction}
	d{\cal{V}}=\left(1-\frac{r_s}{r}\right)^{-1/2}drr^2 sin\theta d\theta d\phi,
\end{equation}
which is only by the factor $\left(1-\frac{r}{r_s}\right)^{-1/2}$ greater than $dV=drr^2sin\theta d\theta d\phi$ in the remote observer coordinates. At $r^*=1.5r_s$ this factor is ca. 1.7, which gives the reduction of $p_F$ caused by gravitational curvature by factor ca. $1.7^{-1/3}\simeq0.84$, which does not change orders in the above estimations. The change of $p_F$ by one order of the magnitude would need the closer approaching the Schwarzschild horizon, at $r\simeq 1.000001 r_s$, i.e., rather distant from the critical $r^*=1.5r_s$. Hence, for the rough estimation of the effect of Fermi sphere collapse the correction (\ref{correction}) is unimportant and can be included as the factor $0.84$ to the right-hand side of Eq. (\ref{fmm}), which does not change the orders in the energy estimation.

In the case when the accretion matter flux is not extremal, then the radiation emitted at the collapse of Fermi spheres at passing the photon sphere rim is low intensive and redshifted proportionally to the reduced Fermi energy. In the case of tidal disruption events (TDE) \cite{tidal1,tidal2} for medium size black holes, which do not form accretion discs, the Fermi sphere collapse also contributes to the total luminosity of transients at occasional matter capturing, as for instance AT 2020neh \cite{ggg50000}, the recently observed ca. one year-lasting episode classified as TDE of a main sequence star (with $1.3$ Sun mass) by the central hole of a dwarf galaxy. Host galaxy of AT 2020neh is SDSSJ152120.07+140410.5 (with $z=0.062$) of mass similar to   Great Magellan Cloud. The short rising phase of the related luminosity allowed \cite{ggg50000} for the estimation of the black hole to be of order of $10^5$ Sun masses. The luminosity, maximal of order of $4 \times 10^{36}$ W lasting shorter than a month, was concentrated in UV and optical range in agreement with conventional models of TDE, though with some contribution in X-ray range of $4.5 \times 10^{34}$ W exceeding these models. This can be, however, explained by inclusion of energy released due to Fermi sphere collapse at passing the photon sphere rim by the star debris at TDE with the compression of fermions adjusted to the greatest observed energy of photons (ca. 10 keV \cite{ggg50000}). Using Eqs (\ref{enn}) and (\ref{fmm}) one can estimate the average luminosity of $4.3\times 10^{34}$ W, which is congruent with the observation in X-ray range \cite{ggg50000}.

Note that the described quantum transition terminates quickly after passing the rim of the photon sphere (along the Fermi golden rule) in agreement with observations, what was, however, unclear for conventional radiation mechanisms from accretion disc, as the falling of the matter across the photon sphere onto the event horizon takes for a remote observer an infinitely long time.

\appendix

\acknowledgments
The author thanks to Professors Lucjan Jacak, Karol Wysoki\'nski, J\'ozef Spa{\l}ek and Andrzej Radosz for discussion and comments on topology and quantum statistics in specific conditions.

\providecommand{\href}[2]{#2}\begingroup\raggedright\endgroup

\end{document}